\documentclass[arguments]{aastex631}

\usepackage{epsfig}
\usepackage{alltt,color}
\usepackage{graphicx}
\usepackage{epstopdf}
\usepackage{float}
\usepackage{palatino}
\usepackage{amsmath}
\usepackage{cases}
\usepackage{pifont}
\usepackage{txfonts}
\usepackage{multirow}
\usepackage{appendix}

\DeclareMathOperator{\erfc}{erfc}

\submitjournal{ApJ}

\shorttitle{}
\shortauthors{}

\begin{document}

\title{Small-scale inhomogeneity effects on coherent solar radio emission}

\correspondingauthor{Xiaowei Zhou}
\email{zhouxw@pmo.ac.cn}

\author[0000-0001-7855-2479]{Xiaowei Zhou}
\affiliation{Purple Mountain Observatory, Chinese Academy of Sciences, Nanjing 210034, People's Republic of China}

\author[0000-0002-3678-8173]{Patricio A. Mu\~{n}oz}
\affiliation{Center for Astronomy and Astrophysics, Technical University of Berlin, Berlin 10623, Germany}

\author[0000-0002-4319-8083]{Jan Ben\'{a}\v{c}ek}
\affiliation{Institute for Physics and Astronomy, University of Potsdam, Potsdam 14476, Germany}

\author[0009-0001-0212-0722]{Lijie Zhang}
\affiliation{Purple Mountain Observatory, Chinese Academy of Sciences, Nanjing 210034, People's Republic of China}
\affiliation{School of Astronomy and Space Science, University of Science and Technology of China, Hefei 230026, People's Republic of China}

\author[0000-0003-2418-5508]{Dejin Wu}
\affiliation{Purple Mountain Observatory, Chinese Academy of Sciences, Nanjing 210034, People's Republic of China}

\author[0000-0001-8058-2765]{Ling Chen}
\affiliation{Purple Mountain Observatory, Chinese Academy of Sciences, Nanjing 210034, People's Republic of China}

\author[0000-0002-9893-4711]{Zongjun Ning}
\affiliation{Purple Mountain Observatory, Chinese Academy of Sciences, Nanjing 210034, People's Republic of China}

\author[0000-0002-5700-987X]{J\"{o}rg B\"{u}chner}
\affiliation{Center for Astronomy and Astrophysics, Technical University of Berlin, Berlin 10623, Germany}
\affiliation{Max Planck Institute for Solar System Research, G\"{o}ttingen 37077, Germany}



\begin{abstract}
Coherent radio emission mechanism of solar radio bursts is one of the most complicated and controversial topics in solar physics.
To clarify the mechanism(s) of different types of solar radio bursts, (radio) wave excitation by energetic electrons in homogeneous plasmas has been widely
studied via particle-in-cell (PIC) code numerical simulations.
The solar corona is, however, inhomogeneous over almost all spatial scales.
Inhomogeneities of the plasma could influence the emission properties of solar radio bursts.
In this paper, we, hence, investigate effects of inhomogeneity (in the magnetic field, plasma density and temperature) of plasmas
in the solar corona on radio wave emission by ring-beam distributed energetic electrons utilizing 2.5-dimensional PIC simulations.
Both the beam and electron cyclotron maser (ECM) instabilities could be triggered
with the presence of the energetic ring-beam electrons.
The resultant spectrum of the excited electromagnetic waves presents a zebra-stripe pattern in the frequency space.
The inhomogeneous density or temperature in plasmas influences the frequency bandwidth and location of these excited waves.
Our results can, hence, help to diagnose the plasma properties at the emission sites of solar radio bursts. Applications of our results
to the solar radio bursts with zebra-stripe pattern are discussed.
\end{abstract}

\keywords{Solar corona; Solar coronal radio emission; Radio bursts}


\section{Introduction}
\label{Introduction}

Radio emissions from the Sun are not only produced by incoherent processes of energetic electrons, but also via
coherent processes involving kinetic instabilities of wave-particle, wave-wave interactions~\citep{Melrose_1991ARA&A..29...31M, Melrose_2017RvMPP...1....5M}.
Solar radio emission is, hence, the most suitable object to diagnose the physical state, acceleration and propagation processes
of high-energy electrons in plasmas of solar activity. The solar radio emission mechanism, especially the coherent radio emission mechanism of
solar radio bursts~\citep{Dulk_1985ARA&A..23..169D}, is one of the most complicated and controversial topics in the solar physics. There has been a constant
controversy between the two types of coherent emission mechanisms, i.e.,
"the plasma emission"~\citep{Ginzburg&Zhelezniakov_1958SvA.....2..653G, Melrose_1970AuJPh..23..871M, Melrose_1970AuJPh..23..885M, Zheleznyakov&Zaitsev_1970AZh....47...60Z, Zheleznyakov&Zaitsev_1970SvA....14..250Z}
and "the electron cyclotron maser (ECM) emission"~\citep{Twiss_1958AuJPh..11..564T, Schneider_1959PhRvL...2..504S, Gaponov59},
for the mechanism of solar radio bursts since a long time. For reviews, see, e.g.,~\citealp{Aschwanden2005psci.book.....A, Melrose_2017RvMPP...1....5M}.

Due to the kinetic nature of coherent radio emissions, fully kinetic Particle-in-Cell (PIC) numerical simulation has been widely applied to figure out mechanism of
coherent radio emission from different space plasmas. For instance,~\citealp{Kasaba_etal_2001JGR...10618693K, Rhee_etal_2009ApJ...694..618R, Umeda_2010JGRA..115.1204U, Thurgood_2015A&A...584A..83T,
Henri_etal_2019JGRA..124.1475H, Chen_etal_2022ApJ...924L..34C, Ni_etal_2020ApJ...891L..25N} investigated the excitation process of electromagnetic waves
by the plasma emission mechanism in weakly magnetized plasmas with $\omega_{ce} < \omega_{pe}$, where $\omega_{ce}$ and $\omega_{pe}$ are the electron gyrofrequency and electron plasma frequency, respectively.
The ECM emission properties in strongly magnetized plasmas with $\omega_{ce} > \omega_{pe}$ were explored by ~\citealp{Pritchett_1984JGR....89.8957P, Lee_etal_10.1063/1.3626562,
Ning_etal_2021A&A...651A.118N, Zhou_etal_2021ApJ...920..147Z, Yousefzadeh_etal_2022ApJ...932...35Y}.
\citealp{Zhou_etal_2020ApJ...891...92Z, Zhou_etal_2022ApJ...928..115Z} systematically compared the excitation efficiency of electromagnetic waves in
differently magnetized plasmas via PIC simulations.
While all PIC simulations in the above mentioned studies initially utilized homogeneous plasmas,
where the magnetic field ($\vec{B}$), density ($n$) and temperature ($T$) of particles are uniform throughout the whole simulation domain.
There is, however, seldom a plasma immersed in a uniform magnetic field in realistic solar coronal environment.
Additionally, plasmas, as we know, are always turbulent due to a diverse processes, including, e.g. reconnection~\citealp{Vlahos&Cargill_2009tsp..book.....V}.
Turbulence involves an energy cascade from large to small scales, and naturally creates inhomogeneities specially
from large space scales to the smallest kinetic scales.
As mentioned in the studies of~\citealp{Melrose1975SoPh...43...79M} and~\citealp{Winglee&Dulk_1986ApJ...307..808W}, the inhomogeneous nature of
plasmas is quite important as it influences properties of the emission spectrum of radio bursts.
Contribution of the inhomogeneity of background magnetic field as well as density and/or temperature in plasmas
on the properties of the coherent radio emission, hence, needs to be further verified and expanded.

\citealp{Pritchett&Winglee1989JGR....94..129P} and~\citealp{Yao_etal_2021JPlPh..87b9003Y}
investigated the effects of nonuniform magnetic field and inhomogeneous
background plasma density on the ECM and plasma emission processes, respectively, via PIC simulations.
However not only the background plasma was ignored but also the background magnetic field ($\vec{B}=[0, B_{y}, 0]$) did not satisfy
$\nabla \cdot \vec{B} = 0$ in the study of \citealp{Pritchett&Winglee1989JGR....94..129P}.
\citealp{Yao_etal_2021JPlPh..87b9003Y}, however, considered inhomogeneity of the background plasma density in the direction along the background magnetic field only with a localized homogeneous energetic electron
beam. The background magnetic field, background plasma temperature and density, temperature of the (localized) energetic electron beam are uniform in that study.
The setups applied by~\citealp{Pritchett&Winglee1989JGR....94..129P} and~\citealp{Yao_etal_2021JPlPh..87b9003Y} would
lead to deviations from the equilibrium state of plasmas. There should be $\vec{J} \times \vec{B} = \nabla
P$ in a magnetohydrodynamics (MHD) equilibrium plasma, where $\vec{J}$ and $P$ are the current density and pressure in plasma, respectively.
Disequilibrium might trigger other instabilities in plasmas,
which would mislead us about the ECM and plasma emission processes.
There are, however, very few analytical kinetic equilibria of plasmas because of the difficulty of solving exactly the
nonlinear integro-differential system of the Vlasov-Maxwell equations. In this study we, hence, consider a MHD equilibrium initial
setup for further investigation of wave excitation.

To our knowledge, there is still no fine observation to
provide a precise distribution of energetic electron in the solar coronal.
Theoretical analysis and numerical simulation have, however, proved that ring-beam momentum
distribution can be formed in the presence of a quasi-perpendicular shock or
magnetic reconnection~\citealp{Vlahos&Sprangle_1987ApJ...322..463V, Vlahos_1987SoPh..111..155V, Bessho_etal_2014GeoRL..41.8688B, Shuster_etal_2014GeoRL..41.5389S,
Shuster_etal_2015GeoRL..42.2586S}.
Additionally, ring-beam distributed energetic electrons have been applied in our previous study for wave excitation in homogeneous
plasmas~\citep{Zhou_etal_2020ApJ...891...92Z}. Applying ring-beam distribution in this study will
benefit to comparison of wave excitation properties between homogeneous and inhomogeneous plasmas.
Energetic electrons are still considered to follow a ring-beam distribution in this study.

In this paper, we, hence, investigated the wave excitation by ring-beam distributed energetic electrons in
inhomogeneous magnetized equilibrium plasmas, where the disequilibrium introduced by the
inhomogeneous magnetic field is balanced by either inhomogeneous density or inhomogeneous temperature of the background plasma.
2.5-dimensional PIC simulations were utilized for the purpose of this study.
Setup of this inhomogeneous magnetized equilibrium plasmas is shown in Section~\ref{Setup}.
The main results of these simulations are presented in Section~\ref{Results}.
We draw our conclusions and discuss applications of our results to
solar radio bursts with zebra-stripe pattern in Section~\ref{Conclusions}. 

\section{Setup of PIC Simulations}
\label{Setup}

We perform this study with the fully kinetic PIC code ACRONYM~\citep{Kilian_etal_2017_PoP}, a fully relativistic electromagnetic code tuned for
the study of kinetic-scale plasma wave phenomena and interactions in collisionless plasmas
~\citep[see, e.g.,][]{Ganse_etal_2012ApJ...751..145G, Schreiner_etal_2017ApJ...834..161S, Munoz2018PhRvE, Zhou_etal_2020ApJ...891...92Z, Yao_etal_2021JPlPh..87b9003Y}.
We use the code in two spatial dimensions and three dimensions in velocity, electromagnetic fields (i.e., 2.5-dimensional or 2.5D).
All quantities in simulations are solved in Cartesian-coordinate spatial-temporal space with Gaussian CGS units, and all simulations have the same spatial and time resolution.
In particular, the grid cell size is $\Delta x = \Delta y = D_{e}$ (where $D_{e}$ is the Debye length of electrons at the boundaries of the y-axis)
with $N_{x} \times N_{y} = 1024 \times 4096$ grid points in the $x-y$ plane of the 2D simulation box. Periodic boundary condition is applied to the boundaries of each axis.
The timestep ($\Delta t$) in our simulations is determined by the inherent length and timescale requirements in
a fully kinetic PIC code, i.e., the Courant-Friedrichs-Lewy (CFL) condition for the speed of light $c$.

\begin{figure*}
\begin{center}
\includegraphics[width=1.0\textwidth]{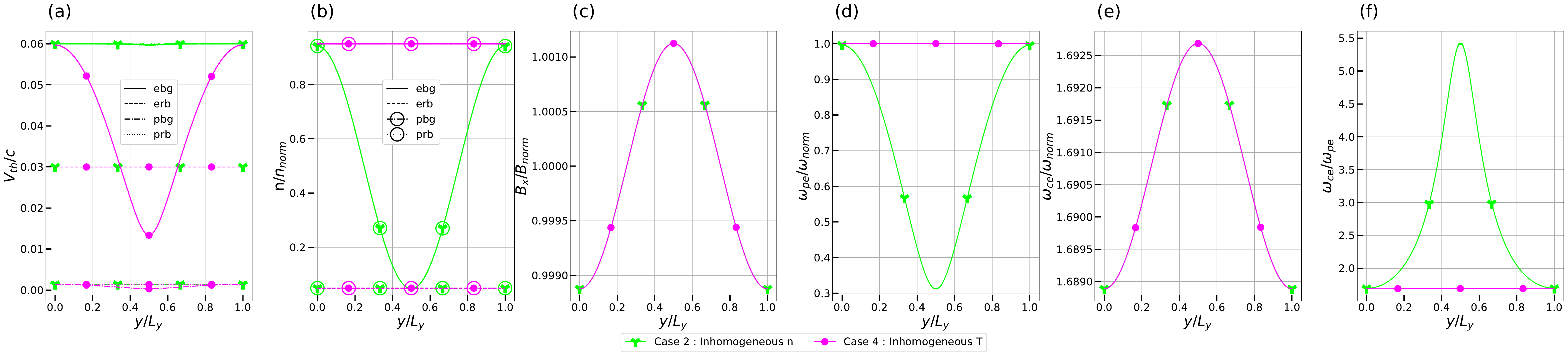}
\caption{Initial y-axis distribution of different (normalized) parameters:
panel (a) and (b) --- thermal velocity and number density of different particle
species. The solid, dashed, dashdot, and dotted lines are separately for background electrons
($ebg$), energetic ring-beam electrons ($erb$), $pbg$-background protons ($pbg$), and $prb$-background protons ($prb$), respectively.
Panel (c) --- magnetic field $B_{x}$,
panel (d) --- electron plasma frequency $\omega_{pe}$,
panel (e) --- electron gyrofrequency $\omega_{ce}$,
panel (f) --- the ratio of the electron gyrofrequency and plasma frequency $\omega_{ce}/\omega_{pe}$.
In each panel, different colors are used to distinguish different cases.
}
\label{Initial_Setup}
\end{center}
\end{figure*}

For our first investigation of the plasma inhomogeneity effects on the wave excitation, a simple inhomogeneous magnetic field following $\vec{B}=[B_{x}(y), 0, 0]$,
with a magnetic field gradient along its perpendicular direction, i.e., the
y-axis, is applied in this study.
And
\begin{equation}
B_{x}(y) = B_{0}\left[\left(1+\displaystyle\frac{B_{e}}{B_{0}}\right)+\left(\eta - 1\right) \cos^{2}\left(\displaystyle\frac{y}{L_{y}}\pi\right)\right]
\label{BB}
\end{equation}
where $L_{y} = N_{y} \Delta y$ is the y-axis size of the simulation domain.
Value of $\eta$ determines the y-axis location of the extremum of $B_{x}(y)$ as well as the total magnetic field strength $|B|$.
The maximum (minimum) $|B|$ is located at the center (boundaries) of the y-axis with $\eta < 1$ ($\eta > 1$), while $\eta = 1$ indicates a uniform magnetic
field throughout the simulation domain. We consider $\eta = 0.1$ as well as $B_{0} = 1.2~G$, $B_{e} = 400 B_{0}$ in this study.
The maximum $|B|$ is about $\sim 480~G$, which is the typical magnetic field strength of
active regions on the Sun~\citep{Aschwanden2005psci.book.....A}. Value of $|B|$ along the y-axis is shown in panel (c) of Fig.~\ref{Initial_Setup}.
And consequently the maxima of the derivatives in Eq.~\eqref{BB} is $\sim 10^{-3}
~G/cm$, which is over four orders of magnitude larger than that of the magnetic field model for active
regions in the Sun, see Eq.(1.4.2) in~\citealp{Aschwanden2005psci.book.....A}, although strength variance of this magnetic
field is only $\sim 1~G$.
Taking account of turbulence and the small space scale covered by PIC simulations, as we mentioned in Sect.\ref{Introduction}, magnetic field with such "huge"
derivative might exist at the small kinetic scales.

For a (MHD) equilibrium plasma environment magnetized by a inhomogeneous magnetic field, there should be
\begin{eqnarray}
\nabla \cdot \vec{B} = 0
\nonumber \\
\vec{J} \times \vec{B} = \nabla P_{th}
\nonumber \\
\nabla \times \vec{B} = \displaystyle\frac{4\pi}{c} \vec{J}
\label{equilibrium_condition}
\end{eqnarray}
where $\vec{J}$ and $P_{th} = \sum_{j} n_{j} k T_{j}$ are the current density and thermal pressure in plasma, respectively. $k$ is the Boltzmann
constant. $n_{j}$ and $T_{j}$ are the number density and temperature of different particle species,
respectively. And $j$ is used to distinguish different particle species.

Four species of particles are employed in each simulation, i.e., energetic electrons ($erb$), background thermal electrons ($ebg$),
$prb$-background protons ($prb$), and $pbg$-background protons ($pbg$), see panel (a) and (b) in
Fig.\ref{Initial_Setup}. $j$ in the above equations can, hence, be $ebg$, or $erb$, or $pbg$, or $prb$.
And there are, hence, $n_{p} = n_{pbg}+n_{prb} = n_{ebg}+n_{erb}$,
where $n_{p}$, $n_{ebg}=n_{pbg}$ and $n_{erb}=n_{prb}$ are the number density of background protons, background electrons and energetic electrons,
respectively, to maintain the global charge neutrality.
For physically realistic results, the proton-to-electron mass ratio has been chosen as $m_{p}/m_{e} = 1836$. And we consider temperature-isotropic plasmas.
Density and temperature of the energetic electrons and $prb$-background protons are initially homogeneous
throughout the whole simulation domain in each simulation.
For an equilibrium plasma environment, where has $P_{B}+P_{th} = constant$, variation of $P_{B}$ (magnetic pressure)
has to be balanced by thermal pressure $P_{th}$ via either inhomogeneous density (Case 2) or inhomogeneous
temperature (Case 4) of both background electrons and $pbg$-background protons.
In other words, density and temperature of the background electrons and $pbg$-background
protons are kept the same in each grid cell of simulation.
For instance, if the density of the background electrons and $pbg$-background protons are nonuniform, their temperatures will be homogeneous (Case 2).
If inhomogeneous temperature is applied to the background electrons and $pbg$-background protons,
they will be homogeneously distributed initially over the whole simulation domain (Case 4).

Momentum ($\vec{u} = \gamma \vec{v}$, where $\gamma = 1/\sqrt{1-v^2/c^2} = \sqrt{1+u^2/c^2}$) of the energetic electrons initially follow a ring-beam distribution
~\citep{Umeda_etal_2007JGRA..112.4212U, Lee_etal_2011PhPl...18i2110L, Kainer_etal_1996JGR...101..495K, Zhou_etal_2020ApJ...891...92Z}, i.e.,
\begin{eqnarray}
F_{rb}(u_{\parallel}, u_{\perp}) = F_{rb\parallel}(u_{\parallel}) F_{rb\perp}(u_{\perp})
\nonumber \\
F_{rb\parallel}(u_{\parallel}) = \displaystyle\frac{1}{\sqrt{2\pi} u_{th\parallel}} \exp\left[-\displaystyle\frac{(u_{\parallel}-u_{d, erb, \parallel})^2}{2 u_{th\parallel}^2}\right]
\nonumber \\
F_{rb\perp}(u_{\perp}) = \displaystyle\frac{1}{2\pi u_{th\perp}^2 A_{\perp}} \exp\left[-\displaystyle\frac{(u_{\perp}-u_{d, erb, \perp})^2}{2 u_{th\perp}^2}\right]
\label{rb_dis}
\end{eqnarray}
where $u_{\parallel}, u_{\perp}$ are the particle momenta along and
perpendicular to the background magnetic field $\vec{B}$, i.e., the parallel and
perpendicular directions are along the x-axis and y-axis, respectively.
Initially we set $u_{th\parallel} = u_{th\perp} = 0.03 c$ and $u_{d, erb, \parallel} = u_{d, erb, \perp} = 0.47 c$, correspondingly the average
initial kinetic energy of energetic ring-beam electrons is $100 keV$ with a Lorentz factor $\gamma_{d, erb} =
1.2$. $100 keV$ is often considered for energetic electrons in solar activity.
While $A_{\perp}$ in Eq.~\eqref{rb_dis} is the normalization constant
\begin{equation}
A_{\perp} = \exp\left[-\displaystyle\frac{u_{rb\perp}^2}{2 u_{th\perp}^2}\right] +
            \sqrt{\displaystyle\frac{\pi}{2}}\displaystyle\frac{u_{rb\perp}}{u_{th\perp}}
            \erfc \left[-\displaystyle\frac{u_{rb\perp}}{\sqrt{2}u_{th\perp}} \right]
\end{equation}

Initial momentum distribution of $prb$-background protons is non-drifting Maxwellian
with a uniform thermal velocity $u_{th,erb} = 0.0014 c$ in each simulation.
Initial momentum distribution of background electrons and $pbg$-background protons are, however, drifting Maxwellian, i.e.,
\begin{align}
f_{ebg/pbg}(u_{x}, u_{y}, u_{z}) = C  \exp\left[-\displaystyle\frac{u_{x,ebg/pbg}^2+u_{y,ebg/pbg}^2+(u_{z,ebg/pbg}-u_{dz,ebg/pbg})^2}{2 u_{th,ebg/pbg}^2}\right]
\label{distribution_3}
\end{align}
where $C$ is the normalization constants. $u_{th,ebg} = 0.06 c$ and $u_{th,pbg} = 0.0014 c$ are the maximal thermal velocity of the background electrons and protons,
respectively, see panel (a) in Fig.~\ref{Initial_Setup}, which indicates that temperature of the background electrons $T_{ebg}$ and protons $T_{pbg}$ are the
same. While $u_{dz,ebg}$ and $u_{dz,pbg}$ are the drift velocity 
(in the out-of-plane direction along z-aixs) of the background electrons and protons. The presence of $u_{dz,ebg}$ and
$u_{dz,pbg}$ are due to the gradient drift of particles~\citep{Zhou_etal_2015ApJ...815....6Z} in inhomogeneous magnetic field Eq.~\eqref{BB}.
There are, hence, $u_{dz,ebg}/T_{ebg} = - u_{dz,pbg}/T_{pbg}$ as well as $\vec{J} = n_{ebg} q_{e} u_{dz,ebg} + n_{pbg} q_{p}
u_{dz,pbg}$, where $q_{e}$ and $q_{p}$ are the charge of electrons and protons, respectively.
The maximum value of $u_{dz,ebg}$ is $1.1\times10^{-4}~c$ and $2.5\times10^{-5}~c$ in Case 2 and Case 4, respectively.
Current due to the drifting motion of the energetic ring-beam electrons along
the background magnetic field is numerically compensated initially to
eliminate effects of large net current on wave excitation~\citep{Melrose_1986islp.book.....M, Matsumoto&Omura1993, Henri_etal_2019JGRA..124.1475H}.

To increase the particle number per cell for the energetic ring-beam electrons, we set the macrofactor of the energetic ring-beam electrons being one quarter of that of background
electrons. At the y-axis boundary of the simulation domain, for example, the number of macroparticles per cell is $950$ for the background electrons and $200$
for the energetic ring-beam electrons, where $n_{erb}/(n_{erb}+n_{ebg})$ is, hence, $0.05$, see panel (b) in Fig.~\ref{Initial_Setup}.
This density ratio was often applied in previous related investigations~\citep[see, e.g.,][]{Lee_etal_2011PhPl...18i2110L, Zhou_etal_2020ApJ...891...92Z}.

Normalization of different parameters are as follows:
$\omega_{norm}$ is the normalization of frequency, which is equal to the maximal $\omega_{pe}$ located at the boundaries of the
y-axis in both plasmas (see panel d of Fig.~\ref{Initial_Setup}). Correspondingly normalization of
the particle number density is $n_{norm} = \omega_{norm}^2 m_{e}/(4 \pi e^{2})$, where $e$ is the charge of electrons.
Time, velocity and distance are normalized by $1/\omega_{norm}$, $c$ and $c/\omega_{norm}$, respectively.
$B_{norm}$, being equal to the mean value of $B_{x}(y)$ in Eq.~\eqref{BB}, is the normalization of the electric and magnetic field strength.
All kinds of energy are normalized by the initial kinetic energy of energetic ring-beam electrons $\varepsilon_{kinetic-ee0}$.

\section{Results}
\label{Results}

\subsection{Energy Evolution}
\label{subsec:Energy_Evolution}

\begin{figure*}
\begin{center}
\includegraphics[width=1.0\textwidth]{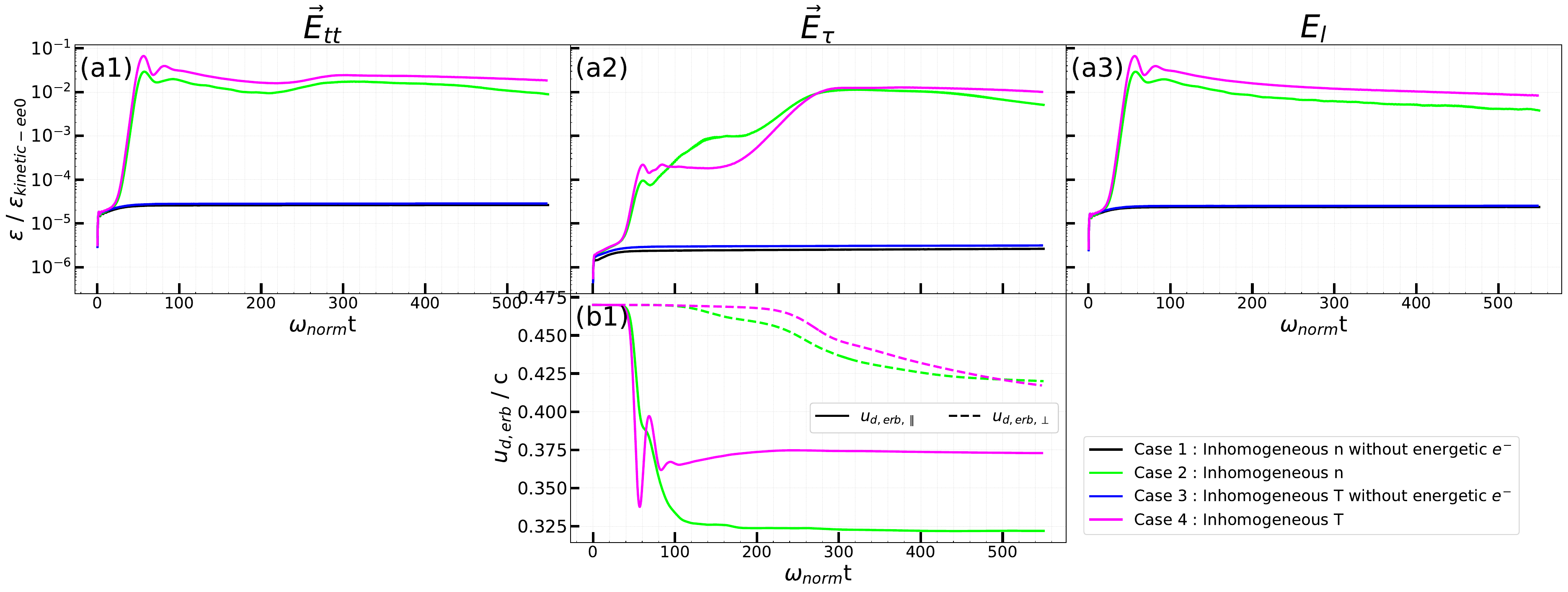}
\caption{Energy evolution of the total ($\vec{E}_{tt}$ in panels a1), transverse ($\vec{E}_{\tau}$ in panel a2) and
         the longitudinal ($E_{l}$ in panel a3) electric fields of waves in the whole simulation domain.
         Panel (b1) presents evolutions of the bulk (or average) drift momenta in the directions along ($u_{d, erb, \parallel}$, solid line) and
         perpendicular ($u_{d, erb, \perp}$, dashed line) to the ambient magnetic field $\vec{B}$ for the energetic ring-beam electrons,
         Different colors in each panel are used to distinguish different cases.
}
\label{Et_El_energy_evolution}
\end{center}
\end{figure*}

Fig.~\ref{Et_El_energy_evolution} shows energy evolution of the total $\vec{E}_{tt} = (E_{x}, E_{y}, E_{z})$ (panel a1),
the transverse $\vec{E}_{\tau} = \vec{E} \times \vec{k}/|\vec{k}|$ (panel a2),
as well as the longitudinal $E_{l} = \vec{E} \cdot \vec{k}/|\vec{k}|$  (panel a3)
electric fields of waves in the simulation domain.
Note that we refer to the properties of the transverse and longitudinal electric fields as those of the electromagnetic and
electrostatic waves, respectively.
Furthermore we also did (test) simulations for thermal plasmas with the same parameters as the above
mentioned plasmas but without energetic electrons (i.e., pure thermal plasmas, Case 1 and Case 3).
With top panels of Fig.~\ref{Et_El_energy_evolution}, one can see that there is almost no electric energy gain in the thermal plasmas, while
the electric energy in plasmas with energetic electrons are more than three orders of magnitude
larger than that in its corresponding thermal plasmas.
These results indicate that the initial setups in this study stay in equilibria as well as there are indeed wave excitations
instead of numerical noise in plasmas with energetic ring-beam electrons.

Panel (b1) of Fig.~\ref{Et_El_energy_evolution} presents evolutions of the bulk
drift momenta in the directions along ($u_{d, erb, \parallel}$) and perpendicular ($u_{d, erb, \perp}$) to the background magnetic field $\vec{B}$ for
the energetic ring-beam electrons.
As we know, the beam and ECM instabilities (leading to the plasma emission and the ECM emission, respectively) are, in general, triggered by
the free energy of energetic ring-beam electrons in the directions along (i.e., $u_{\parallel} \partial F_{rb}(u_{\parallel}, u_{\perp}) / \partial u_{\parallel} > 0$) and
perpendicular (i.e., $\partial F_{rb}(u_{\parallel}, u_{\perp}) / \partial u_{\perp} > 0 $) to the ambient magnetic field $\vec{B}$, respectively~\citep{Melrose_2017RvMPP...1....5M}.
Decrease of $u_{d, erb, \parallel}$ and $u_{d, erb, \perp}$ indicates the free energy release in the corresponding direction.
Evolution of $u_{d, erb, \parallel}$ and $u_{d, erb, \perp}$ can, hence, give us insights on
ideas, e.g., how fast these two instabilities reach their saturations, which instability is more efficient to release its free energy and trigger wave excitation, etc.

For each considered case, comparing panel (a1), (a2) and (a3) of Fig.~\ref{Et_El_energy_evolution}, one can find
that energy gain of the electric field is mainly dominated by the electrostatic
waves, i.e., $E_{l}$ before $\omega_{norm} t = 200$. While after $\omega_{norm} t =
200$, electric energy of the electromagnetic waves ($\vec{E}_{\tau}$) takes another increase and
reaches the same energy level as the electrostatic waves.
Combining the evolution of the parallel and perpendicular bulk momenta for the energetic ring-beam electrons (panel b1 of Fig.~\ref{Et_El_energy_evolution})
with the electric energy evolution of the electrostatic and electromagnetic waves (panel a3 and a2 of Fig.~\ref{Et_El_energy_evolution}, respectively),
one can also discover that the first electric energy enhancement of the electromagnetic waves strongly correlates with the excitation of the electrostatic waves ($E_{l}$)
as well as the decrease of the $u_{d, erb, \parallel}$ before $\omega_{norm} t = 100$. While the second electric energy increase of
the electromagnetic waves occurs together with the decrease of the $u_{d, erb, \perp}$ after $\omega_{norm} t = 200$.
These correspondences indicate the first (second) excitation of the electromagnetic
waves should be attributed to the beam (ECM) instability.
Furthermore, the second saturation (or peak) of the electromagnetic waves is almost two orders of magnitude larger than its first one,
which manifests that the ECM instability can more efficiently excite the electromagnetic waves than the beam instability in plasmas,
same as the previous studies, e.g.,~\citealp{Zhou_etal_2022ApJ...928..115Z}.
Among different cases, energy of both the electrostatic and the electromagnetic
waves are slightly stronger in plasma with inhomogeneous temperature (Case 4) than those in the inhomogeneous-density plasma (Case 2)
during most of the simulation time but beside of the time period $100 < \omega_{norm} t < 260$. Among $100 < \omega_{norm} t < 260$, the stronger energy of the electromagnetic waves
presenting in the inhomogeneous-density plasma (Case 2) is due to its earlier excitation onset of the electromagnetic waves by the ECM instability,
indicated by the decrease of the $u_{d, erb, \perp}$.
%

\subsection{Properties of excited waves}
\label{subsec:section2}

\begin{figure*}
\begin{center}
\includegraphics[width=1.0\textwidth]{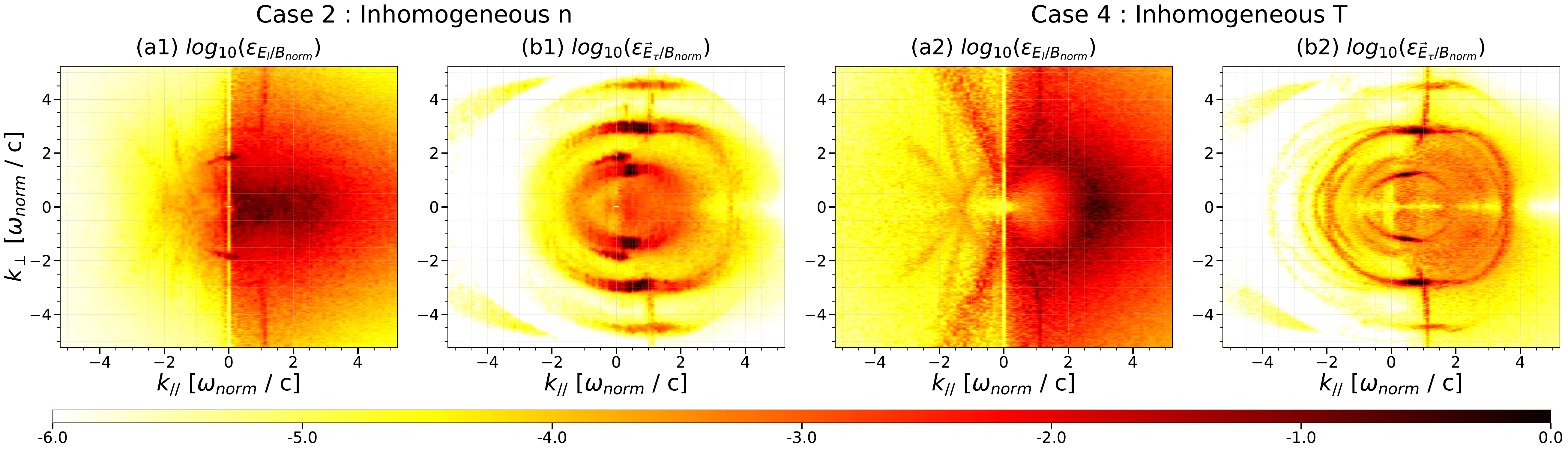}
\caption{$k_{\parallel}-k_{\perp}$ spectrum for the electric field of the electrostatic ($E_{l}$, panels a1 and a2) and electromagnetic $\vec{E}_{\tau}$, panels b1 and b2)
waves over the whole simulation period ($\omega_{norm} t = 0 \sim 546$) in plasmas with inhomogeneous density (Case 2, panels a1 and b1) or inhomogeneous temperature
(Case 4, panels a2 and b2).
}
\label{k_spectrum}
\end{center}
\end{figure*}

Fig.~\ref{k_spectrum} presents the electric energy distribution in the $k_{\parallel}-k_{\perp}$ space of the electrostatic ($E_{l}$, panels a1 and a2)
and electromagnetic $\vec{E}_{\tau}$, panels b1 and b2) waves in plasmas with inhomogeneous density (Case 2, panels a1 and b1) or inhomogeneous temperature
(Case 4, panels a2 and b2).
Due to the symmetry in the direction perpendicular to the ambient magnetic field $\vec{B}$, $k_{\parallel}-k_{\perp}$ spectrum of both the electrostatic and
electromagnetic waves exhibit a symmetry in the $k_{\perp}$ direction. We would, hence, concentrate on the waves with $k_{\perp} > 0$ in the following
investigations.

As one can see that there are big differences in the $k_{\parallel}-k_{\perp}$ spectrum of both the electrostatic and
electromagnetic waves between the inhomogeneous-density and inhomogeneous-temperature plasmas.
For the electrostatic waves, their energies are mainly located at $k_{\parallel} < 3 \omega_{norm}/c$ ($k_{\parallel} > 2 \omega_{norm}/c$) along the
quasi-parallel direction to the background magnetic field in the $k_{\parallel}-k_{\perp}$ space for the inhomogeneous-density (inhomogeneous-temperature) plasma.
The strongly excited electromagnetic waves quasi-perpendicular propagate in both plasmas, but the bandwidth of these electromagnetic waves have wider spreads
in the $k_{\parallel}-k_{\perp}$ space for plasma with inhomogeneous density than those in the inhomogeneous-temperature plasma.
These differences indicate that the excited (both electromagnetic and electrostatic) waves
might have different properties between the inhomogeneous-density and inhomogeneous-temperature plasmas although free energy provided by energetic
ring-beam electrons are the same in these inhomogeneous plasmas.
In the following, we will investigate the properties of these excited waves, e.g., mode of these these excited waves, polarization.

\subsubsection{Electromagnetic Waves}
\label{subsec:section22}

\begin{figure*}
\begin{center}
\includegraphics[width=1.0\textwidth]{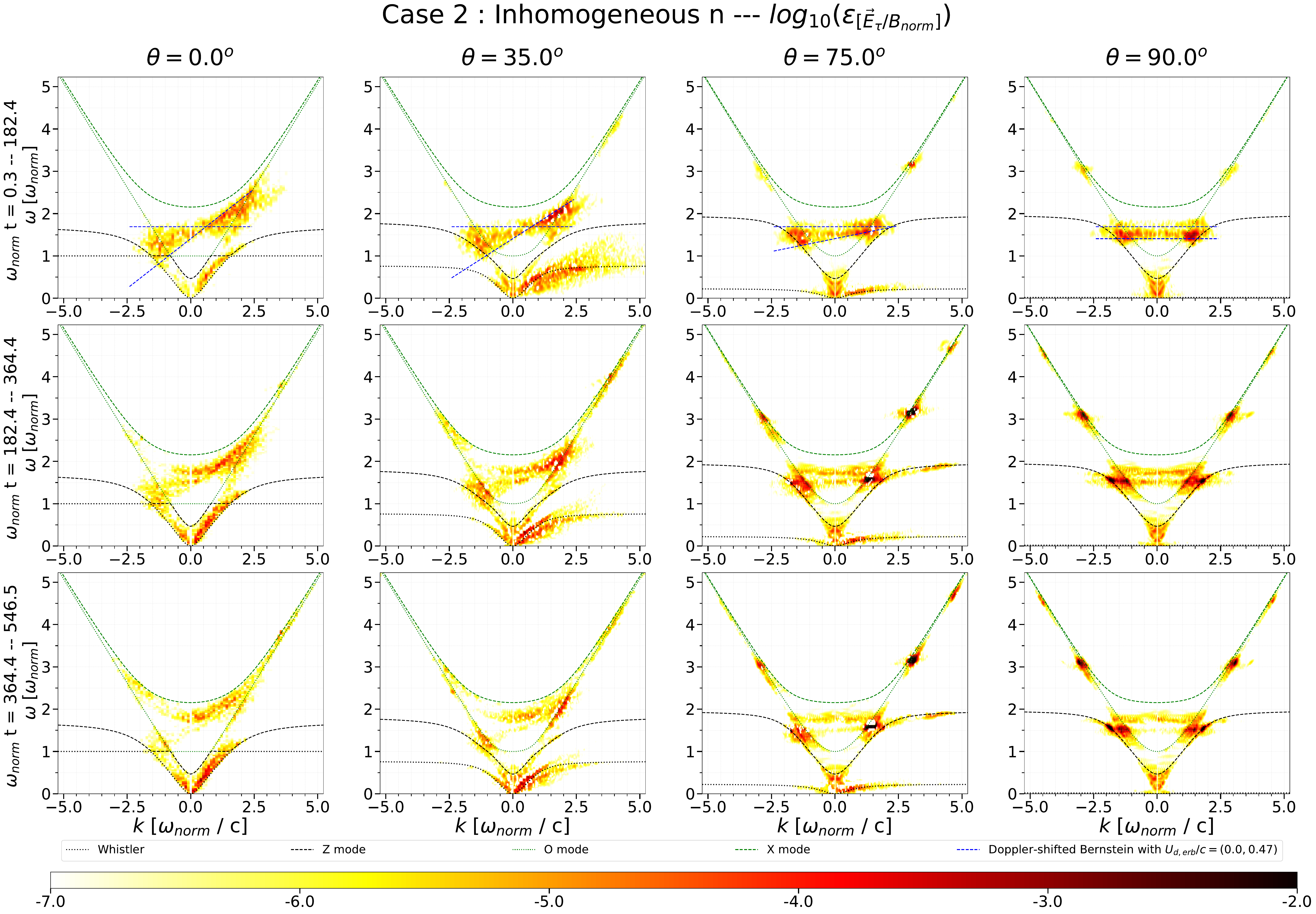}
\caption{$\vec{k} - \omega$ (or dispersion relation) power spectrum of the transverse electric field $E_{\tau}|$
         for the electromagnetic waves propagating along different directions $\theta = |\arctan(k_{\perp}/k_{\parallel})|$ (different columns)
         over the entire space domain but three different time periods (different rows) in the inhomogeneous-density plasma (i.e., Case 2).
         In these panels, $|\vec{k}|=\sqrt{k_{\parallel}^{2}+k_{\perp}^{2}}$ and the sign of $k$ is the same as its parallel component $k_{\parallel}$.
         While the overplotted lines are the four magnetoionic modes in the magnetized cold plasma limit~\citep{Willes&Cairns_2000PhPl....7.3167W}, from bottom to top, they are
         the whistler (black dotted lines), Z (black dashed lines), O (green dotted lines) and X (green dashed lines) modes,
         respectively, where the mean value of $\omega_{ce}$ over the whole simulation domain, i.e. $\omega_{ce,mean}/\omega_{norm} =
         1.69$ and $\omega_{norm}$ for the plasma frequency are applied to the dispersion relation of these four magnetoionic modes.
         Moreover, blue dashed lines around $\omega_{ce,mean}$ in the top
         panels are the Doppler-shifted Bernstein mode with $\omega = \omega_{ce,mean}/\gamma_{d,erb} + \vec{k} \cdot \vec{U_{d,erb}}$,
         where $U_{d,erb}$ is the parallel drift velocity of energetic electrons. $U_{d,erb}$ could range from $0$ to $0.47 c$ and
         the Lorentz factor of energetic electrons $\gamma_{d,erb} = 1.0$ to $1.2$.
}
\label{Et_dispersion_spectrum_case_2}
\end{center}
\end{figure*}

\begin{figure*}
\begin{center}
\includegraphics[width=1.0\textwidth]{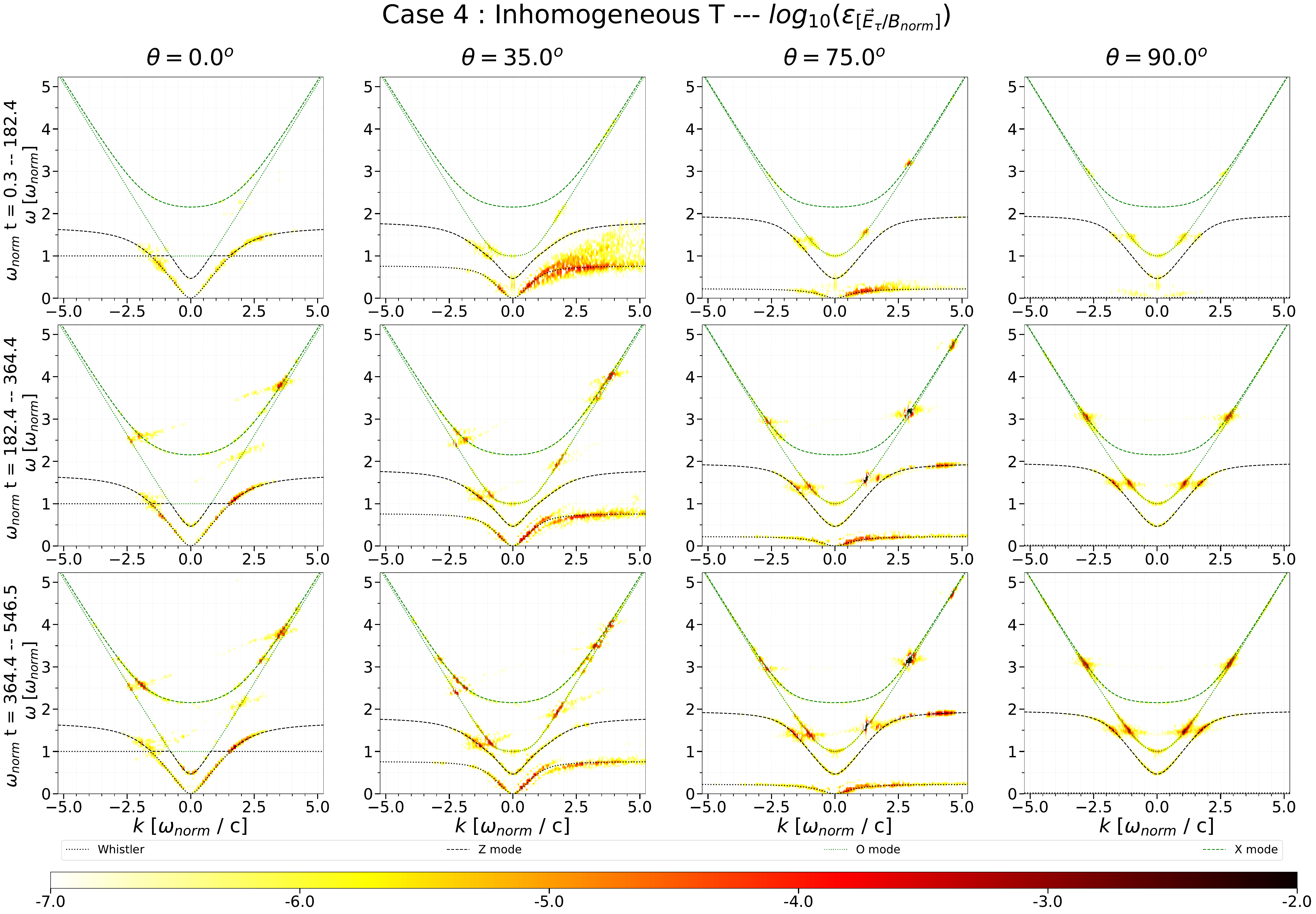}
\caption{Similar to Fig.~\ref{Et_dispersion_spectrum_case_2}, but for the plasma with inhomogeneous temperature (i.e., Case 4).
}
\label{Et_dispersion_spectrum_case_4}
\end{center}
\end{figure*}

\begin{figure*}
\begin{center}
\includegraphics[width=1.0\textwidth]{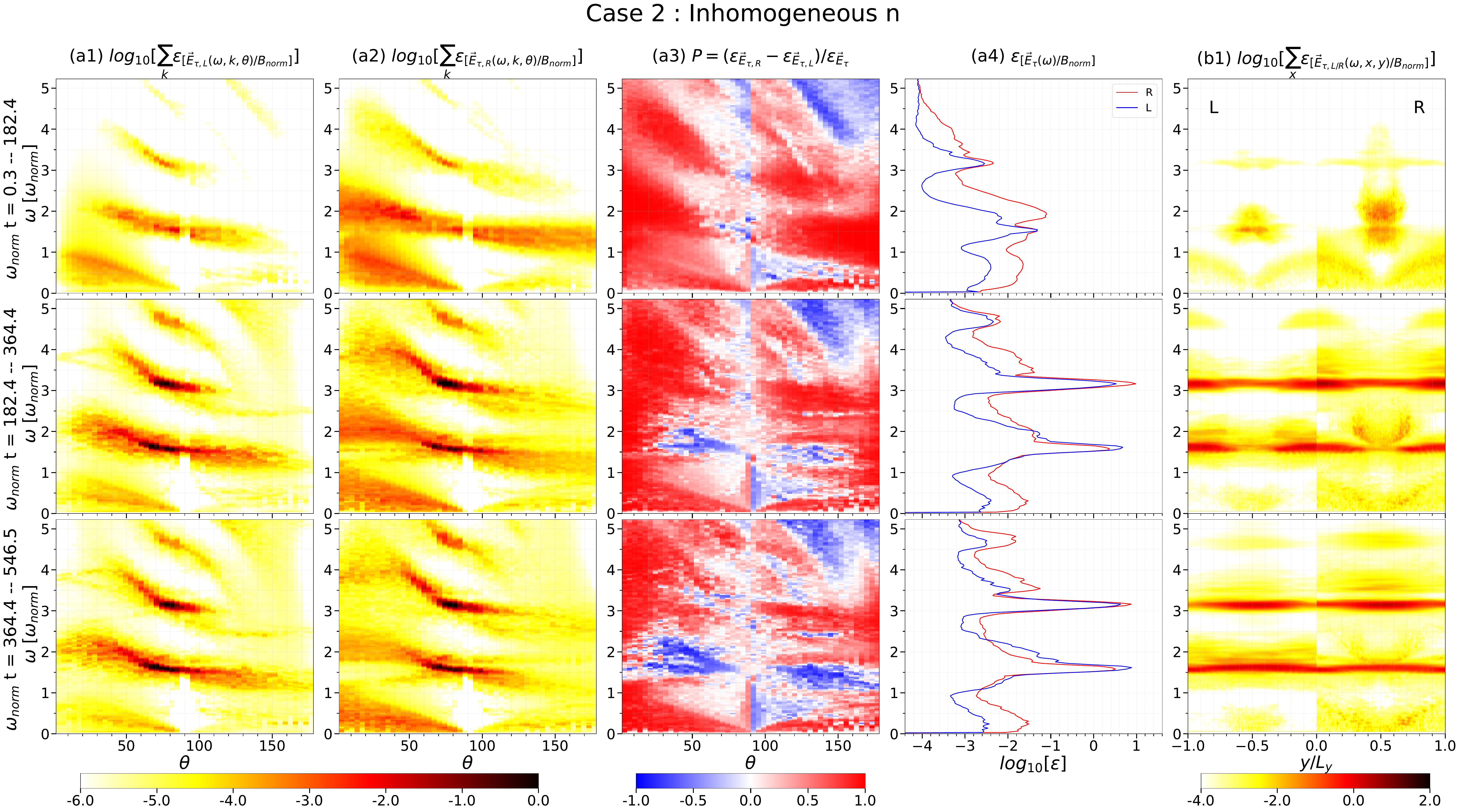}
\caption{For the inhomogeneous-density plasma, columns (a1) and (a2) show the energy distribution for the left-handed ($\epsilon_{\vec{E}_{\tau, L}}$) and right-handed ($\epsilon_{\vec{E}_{\tau, R}}$)
         polarized electric field of electromagnetic waves in the propagating angle-frequency ($\theta - \omega$) space, respectively.
         Polarization degree ($P = (\epsilon_{\vec{E}_{\tau, R}}-\epsilon_{\vec{E}_{\tau, L}})/\epsilon_{\vec{E}_{\tau}}$, where $\epsilon_{\vec{E}_{\tau}} =
         \epsilon_{\vec{E}_{\tau, R}}+\epsilon_{\vec{E}_{\tau,L}}$) of the electromagnetic waves in the whole simulation domain is presented in column (a3).
         Column (a4) displays the energy distribution for the right-handed (red solid line) and left-handed (blue solid line) polarized electric field of
         electromagnetic waves in the frequency $\omega$ space.
         Column (b1) exhibits the energy distribution for the left-handed (right-handed) polarized electric field of electromagnetic waves in the $y-\omega$ space
         with $y/L_{y}<0$ ($y/L_{y}>0$).
         Different rows are used to present evolution of the above-mentioned parameters over three different time periods.
}
\label{Et_polarization_case_2}
\end{center}
\end{figure*}

\begin{figure*}
\begin{center}
\includegraphics[width=1.0\textwidth]{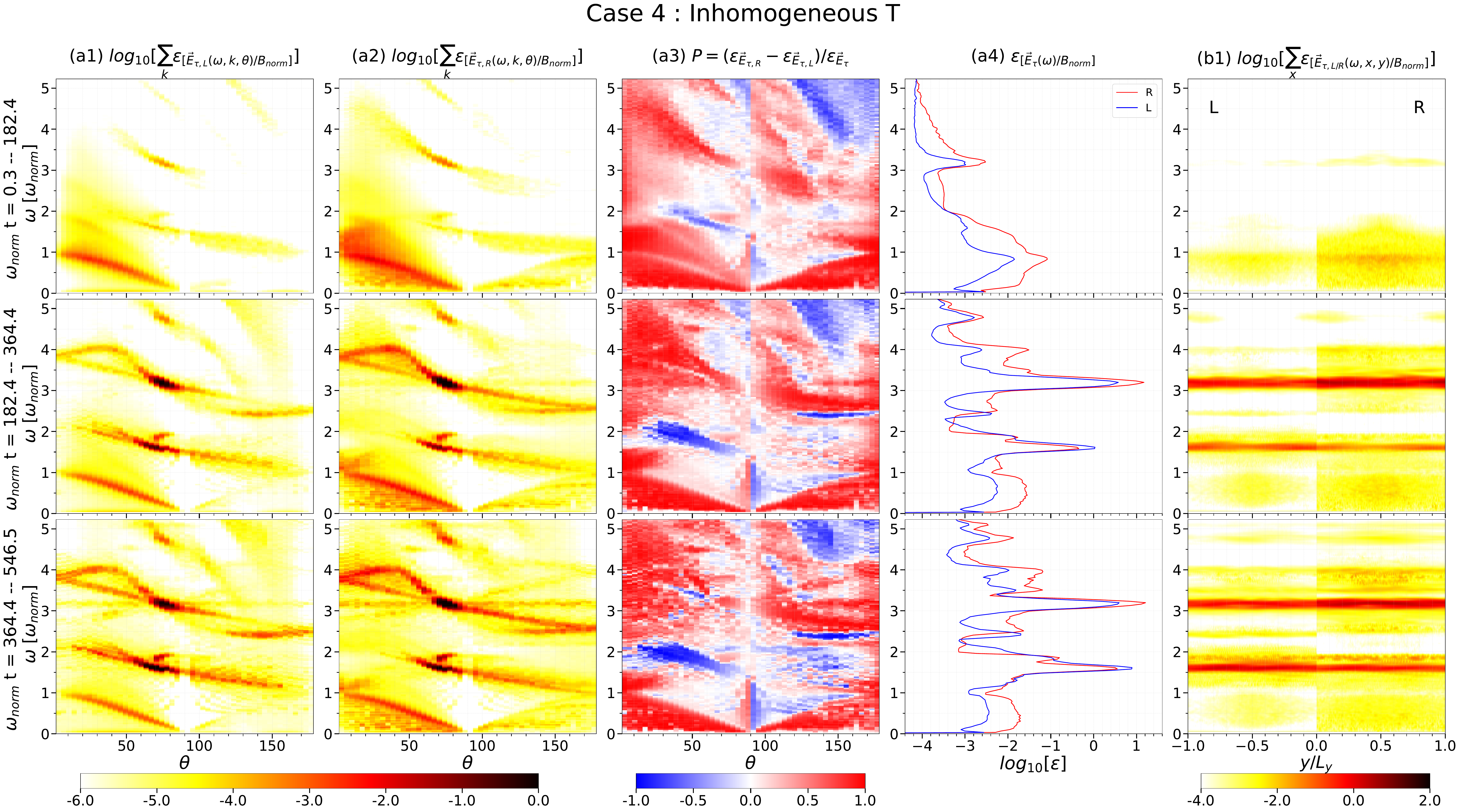}
\caption{Same as Fig.~\ref{Et_polarization_case_2}, but for the plasma with inhomogeneous temperature (i.e., Case 4).
}
\label{Et_polarization_case_4}
\end{center}
\end{figure*}

Figs.~\ref{Et_dispersion_spectrum_case_2} and \ref{Et_dispersion_spectrum_case_4}
present the dispersion spectrum of the electromagnetic waves propagating along different directions over the whole simulation domain.
Polarization property of these electromagnetic waves are shown in the columns (a1), (a2) and (a3) of Figs.~\ref{Et_polarization_case_2} and
\ref{Et_polarization_case_4}.
Spectra of the dispersion relation (Figs.~\ref{Et_dispersion_spectrum_case_2} and \ref{Et_dispersion_spectrum_case_4}) together with
the polarization (Figs.~\ref{Et_polarization_case_2} and \ref{Et_polarization_case_4}) can help us to figure out which wave
mode got stronger excitation.
Consistent with Figs.~\ref{Et_El_energy_evolution} and \ref{k_spectrum}, one can also find that energy of the electromagnetic
waves reaches its peak after $\omega_{norm} t = 180$ and that more strongly excited electromagnetic waves quasi-perpendicular propagate in both plasmas 
with Figs.~\ref{Et_dispersion_spectrum_case_2}, \ref{Et_dispersion_spectrum_case_4}, \ref{Et_polarization_case_2} and \ref{Et_polarization_case_4}.

When $\omega_{norm} t < 182$,
both the the beam and ECM instabilities could be triggered (since $\omega_{norm} t = 100$) processes play roles in the excitation of the electromagnetic waves in the inhomogeneous-density
plasma, see panel (b1) of Fig.\ref{Et_El_energy_evolution}. Correspondingly, the more strongly
excited electromagnetic waves are the right-handed polarized whistler (with $\omega/\omega_{norm} < 1$) and
Doppler-shifted Bernstein or fundamental X-mode, i.e., X1-mode (\citealp{Pritchett_1984JGR....89.8957P} with $\omega/\omega_{norm} \sim \omega_{ce,mean}$) waves,
see top panels in Fig.~\ref{Et_dispersion_spectrum_case_2} and in columns (a1) to (a3) of Fig.~\ref{Et_polarization_case_2}.
$\omega_{ce,mean}$ is the mean value of the electron gyrofrequency $\omega_{ce}$ over the whole simulation domain
and $\omega_{ce,mean} = \overline{\omega_{ce}} = 1.69 \omega_{norm}$, see panel (e) of Fig.~\ref{Initial_Setup}.
Due to evolution of the parallel motion of the energetic ring-beam electrons, these excited Doppler-shifted X1-mode waves cover a quite wide frequency
range especially in the direction along the background magnetic field, see the blue dashed lines covered range in the top panels of Fig.\ref{Et_dispersion_spectrum_case_2}.
Meanwhile energy of the right-handed polarized second harmonic X-mode waves, i.e., X2-mode (with $\omega \sim 2 \omega_{ce,mean}$ and $c|k|/\omega_{norm} \sim 3$)
slightly increase also along the quasi-perpendicular directions during this time period.
In general, for the inhomogeneous-density plasma, the X1-mode waves dominate the energy of all excited electromagnetic waves.
And these excited X1-mode waves can be detected almost over the whole propagating angle $\theta$ space, while the excited whistler and X2-mode waves propagate
mainly with propagating angle $\theta \sim 35^{\circ}$ and $50^{\circ} < \theta < 90^{\circ}$, respectively.
In the inhomogeneous-temperature plasma, there is, however, no obvious excitation for the X1-mode waves,
and instead, most of energy of the excited electromagnetic waves are located in the whistler waves, see top row of Fig.~\ref{Et_dispersion_spectrum_case_4}.
Additionally a weak excitation signal for the left-handed polarized O1-mode waves can be found in the region of $1.5 < \omega/\omega_{norm} <
2$, $c|k|/\omega_{norm} \sim 1$, and $\theta \sim 60^{\circ}$, where one could detect a wave with the left-handed polarization, see top panels in columns (a1) to (a3) of Fig.~\ref{Et_polarization_case_4}.
Excitation of the whistler and O1-mode waves in the inhomogeneous-temperature plasma could be due to the
beam instability (i.e., the plasma emission process) only since there is still no obvious energy decrease in the perpendicular momentum profile of the energetic ring-beam
electrons before $\omega_{norm} t = 200$.

During the time period $\omega_{norm} t = 182 - 364$, when excitation of the electromagnetic waves is taken over by the ECM instability
(see Sec.~\ref{subsec:Energy_Evolution}), the fundamental waves (with $\omega \sim \omega_{ce,mean}$, i.e., the X1 waves,
the right-handed polarized Z-mode, and the left-handed polarized O-mode waves) as well as the second (with $\omega \sim 2 \omega_{ce,mean}$),
and third (with $\omega \sim 3 \omega_{ce,mean}$) harmonic waves get energy enhanced in both plasmas,
see the middle panel in columns (a1) to (a4) of Figs.~\ref{Et_polarization_case_2} and \ref{Et_polarization_case_4} together with the
middle row of Figs.~\ref{Et_dispersion_spectrum_case_2}, \ref{Et_dispersion_spectrum_case_4}, respectively.
Frequency of the fundamental and high order harmonic electromagnetic waves are, hence, related to $\omega_{ce,mean}$ in this study.
Among these excited electromagnetic waves, the more strongly excited ones are located in the second harmonic branch and
propagate mainly quasi-perpendicular with $50^{\circ} < \theta < 90^{\circ}$
(see the middle panel in columns a1, a2 of Figs.~\ref{Et_polarization_case_2} and \ref{Et_polarization_case_4}).
Due to the energy enhancement of the fundamental O-mode waves, there are a few large regions dominated
by the left-handed polarized electromagnetic waves in the $\theta - \omega$
space in both plasmas, e.g., ($30^{\circ} < \theta < 50^{\circ}, 1.5<\omega/\omega_{norm}<2$) in the inhomogeneous-density plasma and
($20^{\circ} < \theta < 90^{\circ}, 1.5<\omega/\omega_{norm}<2$) in the inhomogeneous-temperature plasma, see the middle panel in columns (a3) of
Figs.~\ref{Et_polarization_case_2} and \ref{Et_polarization_case_4}, respectively.
Beside of the above mentioned regions, there is another left-handed polarization dominant region ($120^{\circ} < \theta < 170^{\circ}, \omega/\omega_{norm} \sim
2.5$) located at the second harmonic branch in the inhomogeneous-temperature plasma, even though the backward propagating O-mode waves in this region (see the first and second panels in the
second row of Fig.~\ref{Et_dispersion_spectrum_case_4}) are not so strongly excited as those forward propagating O-mode waves.
Overall, in both plasmas, energy of the fundamental waves is dominated by the O-mode waves and a left-handed polarization could mostly be
detected, while the right-handed polarized X-mode waves are in charge of the properties of the second and third harmonic waves
(see the middle panel in the column a4 of Figs.~\ref{Et_polarization_case_2} and \ref{Et_polarization_case_4}).
Considering a remote detection, left-handed polarization degree of the fundamental branch could be even
lager than that in their source region, since the right-handed polarized Z-mode waves in the fundamental branch could not escape from their source
region if these Z-mode waves do not experience wave mode conversion to escaping electromagnetic waves.

After $\omega_{norm} t = 364$, when excitation of the electromagnetic waves has been reached their saturation (see panel a2 of Fig.~\ref{Et_El_energy_evolution}),
there is no excitation of a new mode wave, and polarization properties of different harmonic branches are similar as those in the earlier time period.
Energy partition among these excited electromagnetic waves, however, changes in each kind of plasma. For
instance, energy of the second harmonic right-handed polarized waves decreases along with a narrower and narrower peak band around $2 \omega_{ce,mean}$,
while the fundamental and third harmonic waves gain energies after $\omega_{norm} t \sim
364$, see the middle and bottom panels in the column (a4) of Figs.~\ref{Et_polarization_case_2} and \ref{Et_polarization_case_4}.
In the inhomogeneous-density plasma, energy of the left-handed polarized fundamental
waves even reach a similar level to that of the second harmonic waves after $\omega_{norm} t = 364$.
There is, hence, a possibility that the second harmonic waves (2H) might decay into the fundamental (F) and third (3H) harmonic
waves, i.e., $2H + 2H \rightarrow F + 3H$ during the nonlinear period of the wave excitation in both plasmas.
Due to the energy enhancement of the O-mode waves in the fundamental branch during this time period, the
left-handed polarization dominated regions get enlarged and enhanced around $\omega_{ce,mean}$ in the $\theta - \omega$
space in both plasmas, see the middle and bottom panels in column (a3) of Figs.~\ref{Et_polarization_case_2} and \ref{Et_polarization_case_4}.

Additionally, note that there are different energy evolutions for the low frequency ($\omega/\omega_{norm} < 1$) whistler waves in these two kinds
of inhomogeneous plasmas.
In the inhomogeneous-density plasma, peak energy of the whistler waves increases and moves to lower frequencies with the evolution of the plasma
system, accompanied by the energy decrease of the X1-mode waves (see Fig.~\ref{Et_dispersion_spectrum_case_2} and column a4 of
Fig.~\ref{Et_polarization_case_2}). The X1-mode waves might, hence, contribute to the energy gain of the whistler waves via wave-wave interaction.
On the contrary, total energy of the whistler waves always decreases after its excitation in the inhomogeneous-temperature plasma (Case
4), where excitation of the X1-mode waves is missing (see Fig.~\ref{Et_dispersion_spectrum_case_4} and column a4 of Fig.~\ref{Et_polarization_case_4}).
Moreover, one can also find that, in the $\theta - \omega$ space with the increase of the wave propagating angle $\theta$,
there is a frequency drift from higher to lower frequencies in each harmonic brand for both inhomogeneous plasmas
(columns a1 and a2 of Figs.~\ref{Et_polarization_case_2} and \ref{Et_polarization_case_4}). Similar to the excited Doppler-shifted X1-mode
in the inhomogeneous-density plasma, this frequency drift might also be due to the Doppler effect due to the parallel drifting motion of the energetic ring-beam
electrons, i.e., $\omega = h \omega_{ce,mean}/\gamma_{d,erb} + k_{\parallel} u_{d,erb,\parallel}
\cos\theta$, where $h$ is the harmonic index. A smaller wave propagating angle
$\theta$ could lead to a larger frequency shift from the characteristic frequency $h \omega_{ce,mean}$ of the h-harmonic branch.
As a result, frequency separation between adjacent harmonic branches will be influenced by the wave propagating angle $\theta$, property of
the dominant electromagnetic waves as well as harmonic index $h$. Equidistant frequency separation, hence, more likely appear among harmonic branches with larger harmonic index
$h$.
Last but not the least, these excited electromagnetic waves have wider frequency
bandwidths in the inhomogeneous-density plasma than those in the inhomogeneous-temperature
plasma (columns a1 and a2 of Figs.~\ref{Et_polarization_case_2} and \ref{Et_polarization_case_4}),
which coincides with the theoretical study of~\citealp{Winglee&Dulk_1986ApJ...307..808W}.
Correspondingly spectrum of the excited electromagnetic waves has a more wider spread over the $k_{\parallel}-k_{\perp}$ spaces in the inhomogeneous-density
plasma (see panels b1 and b2 of Fig.~\ref{k_spectrum}).

Column (b1) of Figs.~\ref{Et_polarization_case_2} and \ref{Et_polarization_case_4} exhibits the energy distribution for both the left-handed and
right-handed polarized electric field of electromagnetic waves in the $y-\omega$ space over different time periods. In order to separate the left-handed
and right-handed polarized waves, electric energy of these left-handed polarized
waves are present with $y/L_{y}<0$, their locations along the y-axis of the simulation domain are equal to $|y|$.
One could, hence, figure out the excitation location (or source region) of these excited electromagnetic waves with these panels.

In the inhomogeneous-density plasma (Case 2), before $\omega_{norm} t \sim 182$, the right-handed elliptically polarized X1-mode waves with $\omega/\omega_{norm} \sim 2$
are excited around the center of the y-axis, where has a quite tenuous background plasma (see panel b of Fig.~\ref{Initial_Setup}).
During the same period, the whistler waves with $\omega/\omega_{norm} < 1$ have more energies located further away from the central y-axis.
In addition, due to the dispersion relation of the whistler wave depending on the local plasma frequency $\omega_{pe}$ (for instance, the resonance frequency of
the whistler wave is $\sqrt{\left(\omega_{pe}^{2}+\omega_{ce}^{2}-\sqrt{(\omega_{pe}^{2}+\omega_{ce}^{2})^{2} - 4\omega_{pe}^{2}\omega_{ce}^{2}\cos ^{2} \theta}\right)/2}$~\citealp{Willes&Cairns_2000PhPl....7.3167W}),
these excited whistler wave cover a wider and wider frequency range from the center to the boundary of the y-axis, i.e., $|y|/L_{y} = 0$ to
$1$.
During the second growth period of the electromagnetic waves, i.e., $\omega_{norm} t = 182 \sim 364$, energy of those stronger excited (fundamental, second, and
third) electromagnetic harmonic waves are also mainly located at the boundaries of the y-axis.
During the nonlinear period $\omega_{norm} t > 364$, most of these stronger excited electromagnetic waves emplace their energy more around the center of the y-axis
due to the propagation of these electromagnetic waves as well as weaker wave damping there with a low plasma density.
Similar location between the energy-decreased X1-mode and energy-enhanced whistler waves at the end of simulation indicates
again that the X1-mode waves might contribute to the energy gain of the whistler waves via wave-wave interactions.

In the inhomogeneous-temperature plasma (Case 4), different from those in the inhomogeneous-density plasma, there is no excitation signal for the X1-mode waves,
and furthermore energy of all excited electromagnetic waves are more or less homogeneously distributed along the y-axis, similar to the previous related studies
in homogeneous plasmas.

Location of the X1-mode source region with low plasma density in the inhomogeneous-density plasma as well as excitation absence of the X1-mode
waves in the inhomogeneous-temperature plasma with denser background plasma
indicate that excitation of the X1-mode waves prefer to occur in low density
plasmas with a larger frequency ratio of $\omega_{ce}/\omega_{pe}$.
High-density source region for other excited electromagnetic (e.g., Z, harmonics of the O, higher harmonic X-mode)
waves implies that stronger excitation of these waves happen in plasmas with a relative smaller frequency ratio of $\omega_{ce}/\omega_{pe}$.
These conclusions are, actually, consistent with the analytical investigations for the growth rate of different wave modes excited by the ECM
instability, e.g.,~\citealp{Tong_etal_2017PhPl...24e2902T, Chen_etal_2017JGRA..122...35C,
Zhao_etal_2016ApJ...822...58Z}. In these investigations, they found that the
maximum growth rate of the X1-mode (other electromagnetic mode) waves located at a region with larger (smaller) frequency ratio of $\omega_{ce}/\omega_{pe}$
in spite of the momentum distribution of the energetic electrons.
Additionally, different source regions of the excited waves between the inhomogeneous-density and
inhomogeneous-temperature plasmas indicate that density of the background plasma has a stronger influence on wave excitation than its temperature.
That might be the reason of neglecting the temperature effect on the growth rate of different wave modes
in the above mentioned analytical investigations for (non-relativistic) space plasmas.

\subsubsection{Electrostatic Waves}
\label{subsec:section21}

\begin{figure*}
\begin{center}
\includegraphics[width=1.0\textwidth]{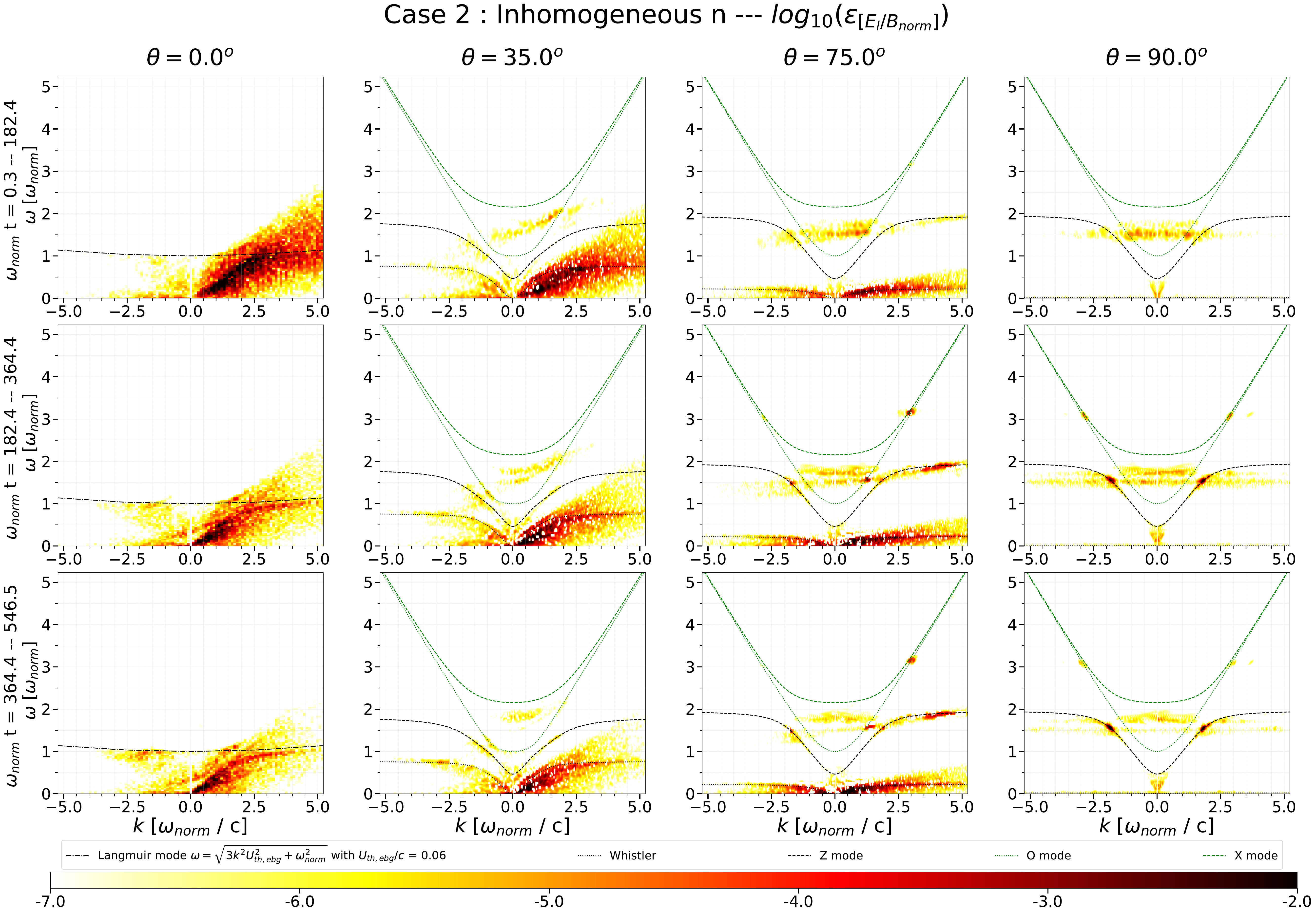}
\caption{Similar to Fig.~\ref{Et_dispersion_spectrum_case_2}, but for the longitudinal electric field $E_{l}|$
         in the inhomogeneous-density plasma (i.e., Case 2).
         In each panel of the left-hand first column, the overplotted dashdot line presents the dispersion relations of the Langmuir wave
         ($\omega = \sqrt{3 k^2 u_{th,ebg}^2 + \omega_{norm}^2}$). The overplotted line in other panels are the same with that in
         Fig.~\ref{Et_dispersion_spectrum_case_2}.
}
\label{El_dispersion_spectrum_case_2}
\end{center}
\end{figure*}

\begin{figure*}
\begin{center}
\includegraphics[width=1.0\textwidth]{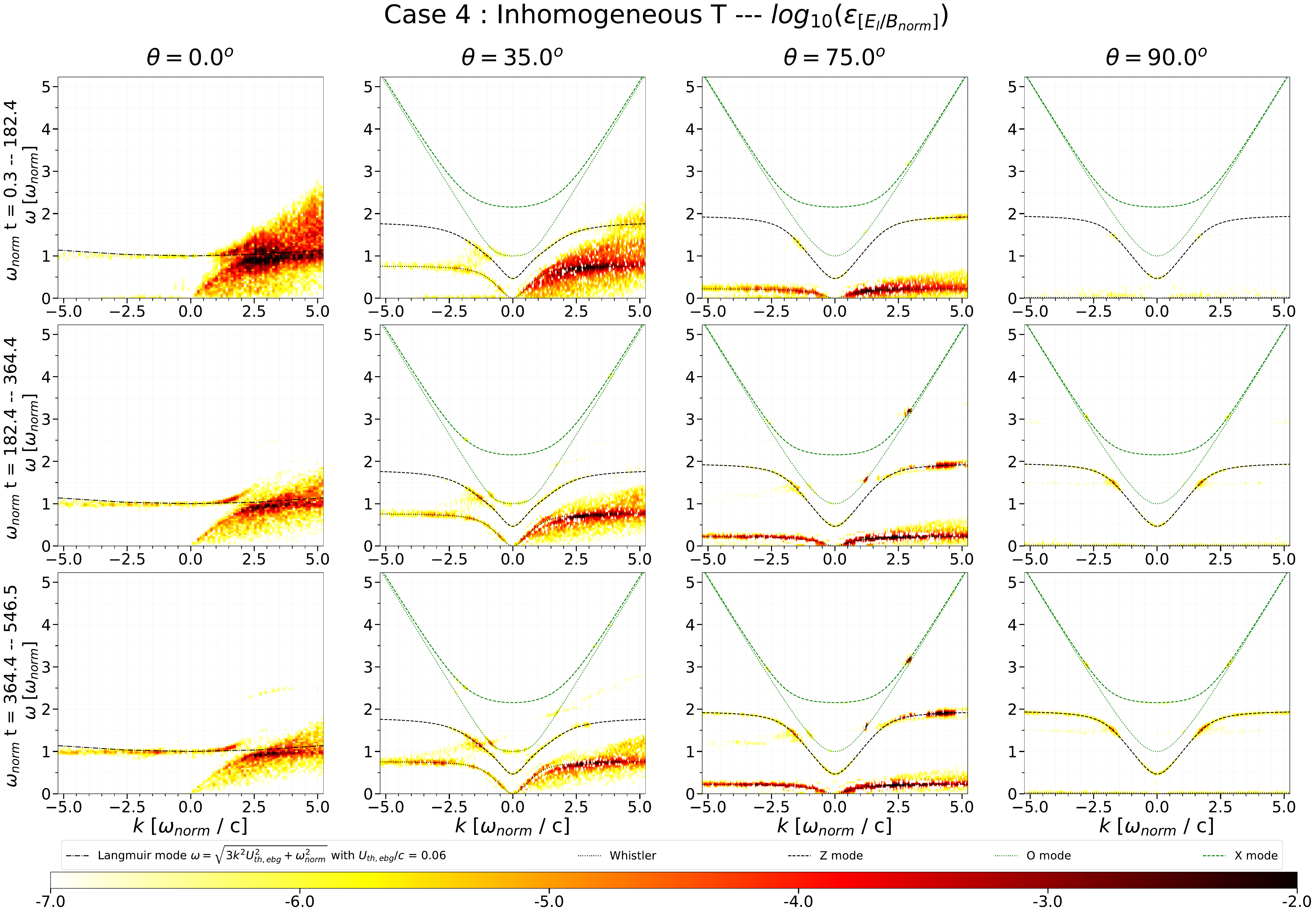}
\caption{Same as Fig.~\ref{El_dispersion_spectrum_case_2}, but for the plasma with inhomogeneous temperature (i.e., Case 4).
}
\label{El_dispersion_spectrum_case_4}
\end{center}
\end{figure*}

\begin{figure*}
\begin{center}
\includegraphics[width=1.0\textwidth]{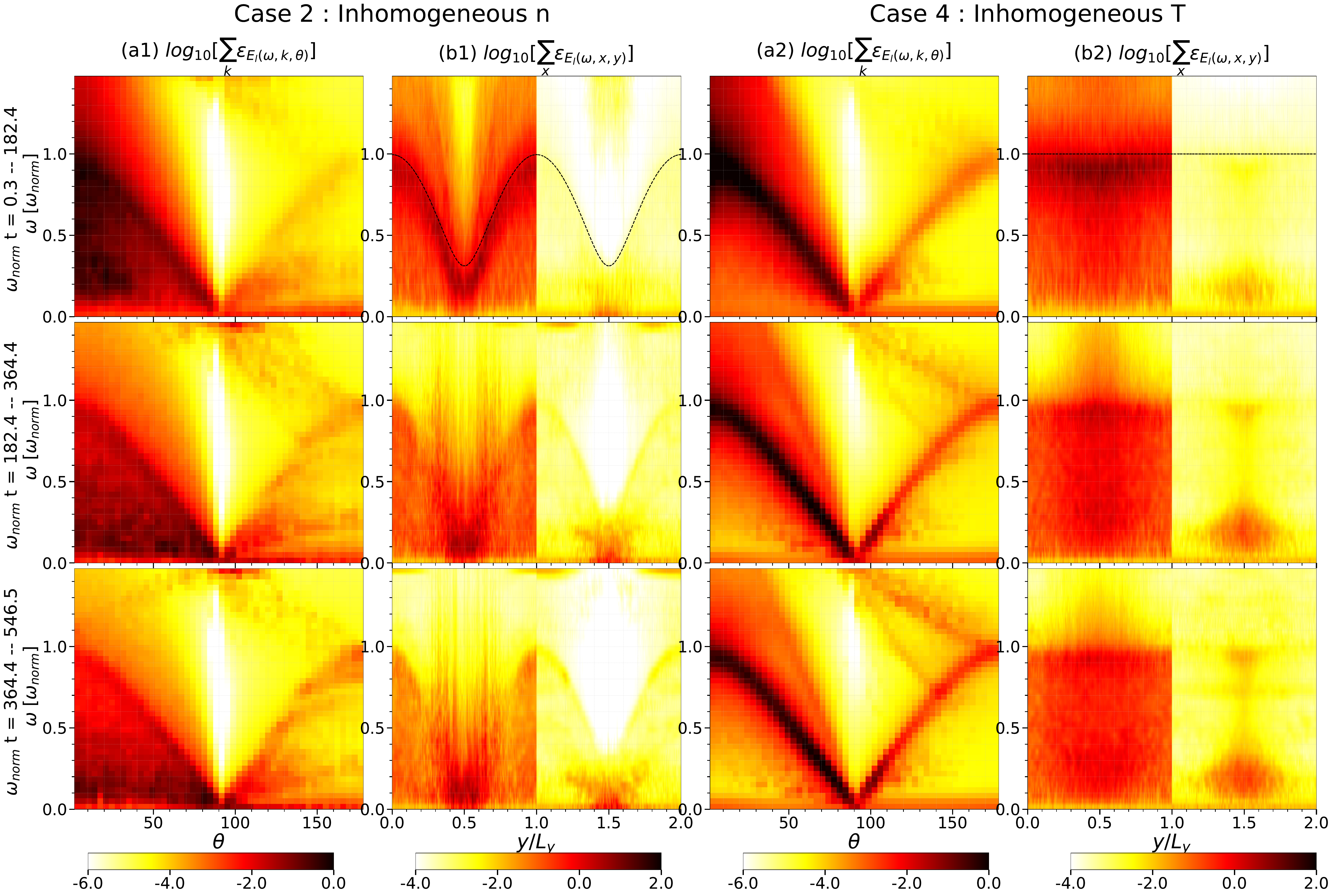}
\caption{Energy distribution of the longitudinal electric field $E_{l}$ in the propagating angle-frequency ($\theta - \omega$, presented in columns a1, a2)
         as well as y-axis-frequency spaces (presented in columns b1, b2) for the inhomogeneous-density (Case 2) and inhomogeneous-temperature (Case 4) plasmas, respectively.
         In columns (b1) and (b2), region with $y/L_{y}<1$ ($y/L_{y}>1$) is for the longitudinal electric field $E_{l}$ propagating with
         $k_{\parallel} > 0$ ($k_{\parallel} < 0$). And the overplotted dashed line in the top panel of columns (b1) and (b2) present the
         localized plasma frequency $\omega_{pe}$ along the y-axis, same with panel (d) in Fig.~\ref{Initial_Setup}.
         While different rows are used to present evolution of the above-mentioned parameters over three different time periods.
}
\label{El_in_other_spaces}
\end{center}
\end{figure*}

Figs.~\ref{El_dispersion_spectrum_case_2}, \ref{El_dispersion_spectrum_case_4} and \ref{El_in_other_spaces} present
energy distribution of the longitudinal electric field $E_{l}$ in the dispersion ($k - \omega$), propagating angle-frequency ($\theta - \omega$)
as well as y-axis-frequency spaces. Longitudinal electric field $E_{l}$ can, to some extent, exhibits properties of electrostatic waves
although some electromagnetic waves contain longitudinal electric field also, e.g., the whistler, Z and X-mode waves~\citep{Willes&Cairns_2000PhPl....7.3167W, Huang&Lyu2019PhPl...26i2102H}.

As we know that two kinds of electrostatic waves can be excited in the frequency range of $\omega \gg \omega_{ci}$ in the presence of an electron beam in
plasma, i.e., the Langmuir wave and the beam wave $\omega = k_{\parallel} v_{b}$, where $\omega_{ci}$ and $v_{b}$ are the gyrofrequency of
protons and the bulk velocity of the energetic electron beam, respectively~\citep{Gary_1993tspm.book.....G}.
With Figs.~\ref{El_dispersion_spectrum_case_2}, \ref{El_dispersion_spectrum_case_4} and columns (a1) and (a2) of
Fig.~\ref{El_in_other_spaces}, one can find that most energy of these excited electrostatic waves are located in those quasiparallel
(in particular parallel) propagating ones with $\omega/\omega_{norm} < 2$, and we consider that these parallel propagating and strongly excited electrostatic waves are the
Langmuir wave in both plasmas.
That can be easily identified in the inhomogeneous-temperature plasma (Case 4) with uniform background electron density and plasma
frequency (see panels b and d of Fig.~\ref{Initial_Setup}), where one can find that the mostly excited parallel propagating electrostatic waves are located near
the uniform Langmuir wave in the dispersion relation space (see the left-hand first column of Fig.~\ref{El_dispersion_spectrum_case_4}).
In the inhomogeneous-density plasma (Case 2), with top panel in column (b1) of Fig.~\ref{El_in_other_spaces}, one can find that frequency of these stronger excited
electrostatic waves also have similar y-axis dependence to the background inhomogeneous plasma frequency.
The beam-mode wave would be, however, homogeneously distributed along the y-axis if the corresponding energetic electron beams are
uniformly distributed in plasma, which we apply in this study.
Due to the wider range of the plasma frequency $\omega_{pe}$ in the inhomogeneous-density plasma, spectrum of the excited electrostatic Langmuir wave have a more wider
spread in the $k-\omega$ (see the left-hand column of Fig.~\ref{El_dispersion_spectrum_case_2}) as well as the $k_{\parallel}-k_{\perp}$ spaces
(see panel a1 of Fig.~\ref{k_spectrum}).

With the increase of the propagating angle $\theta$, in both inhomogeneous plasmas, the dispersion relation of these more energized longitudinal electric fields $E_{l}$ coincide well
with the dispersion relation of the magnetoionic modes in the magnetized cold plasma limit as well as the transverse
electric fields $E_{\tau}$ of those excited electromagnetic waves (i.e., Figs.~\ref{Et_dispersion_spectrum_case_2} and \ref{Et_dispersion_spectrum_case_4}).
That indicates that the longitudinal electric fields $E_{l}$ with larger $\theta$ are the longitudinal part of those excited electromagnetic
waves, and these excited (pure) electrostatic waves are mainly the Langmuir waves with $\omega/\omega_{norm} \leq 1$ propagating quasiparallel
to the background magnetic field.
In the followings, we, hence, consider only the $E_{l}$ with $\omega/\omega_{norm} < 1.5$ for electrostatic Langmuir waves in both inhomogeneous plasmas in Fig.~\ref{El_in_other_spaces},
where the forward (with $k_{\parallel} > 0$) and backward (with $k_{\parallel} < 0$) propagating electrostatic waves are separated
with $(\theta < 90^{\circ}, \theta > 90^{\circ})$ in columns (a1, a2) and $(y/L_{y}<1, y/L_{y}>1)$ in columns (b1, b2), respectively.

As time goes on, these strongly excited forward (quasiparallel) propagating Langmuir waves, however, dissipate faster than the longitudinal electric fields $E_{l}$
of those quasiperpendicular propagating electromagnetic waves, see Figs.~\ref{El_dispersion_spectrum_case_2} and \ref{El_dispersion_spectrum_case_4}, probably due to the
strong Landau damping processes of Langmuir waves.
And this dissipation mainly occurs at the y-axis boundary of the simulation domain in both plasmas
(see the region with $y/L_{y}<1$ in columns b1 and b2 of Fig.~\ref{El_in_other_spaces}),
where the larger density and/or higher temperature of the background electrons (than those around the center of the simulation domain, see panels a and b of
Fig.~\ref{Initial_Setup}) benefit the Landau damping process.
While with the energy decrease of these strongly excited forward propagating Langmuir waves, energy of their corresponding backward propagating parts
increase, see the left column of Figs.~\ref{El_dispersion_spectrum_case_2} and \ref{El_dispersion_spectrum_case_4} as well as
columns (a1) and (a2) of Fig.~\ref{El_in_other_spaces}. That implies that those backward propagating Langmuir waves probably come from scattering
of these forward propagating Langmuir waves. Those backward propagating Langmuir waves, however, have different locations along the y-axis in the inhomogeneous-density
and inhomogeneous-temperature plasmas.
In the inhomogeneous-density plasma, frequency of the Langmuir waves vary along y-axis but the scattered backward propagating Langmuir waves are
mainly located around $\omega/\omega_{norm} \sim 1$ (see the bottom panel in the left column of Fig.~\ref{El_dispersion_spectrum_case_2})
corresponding to the plasma frequency at the y-axis boundary of the simulation domain.
Most energy of these backward propagating Langmuir waves with $\omega/\omega_{norm} \sim 1$ are indeed located at the y-axis boundary,
see the region with $y/L_{y}>1$ in column b1 of Fig.~\ref{El_in_other_spaces}, where the larger density of the background particles would enhance scattering of Langmuir waves.
On the contrary, the backward propagating Langmuir waves in the the inhomogeneous-temperature plasma are mainly located at the center of the plasma system
(see the region with $y/L_{y}>1$ in column b2 of Fig.~\ref{El_in_other_spaces}) due to the weaker dissipation of the forward propagating Langmuir waves there.
Therefore, in total, energy dissipation of these forward propagating Langmuir waves in plasmas are more likely due to both the Landau damping and
the wave scattering processes.

In the plasma emission mechanism, as we know, scattering of Langmuir waves can lead to the enhancement of the backward propagating Langmuir waves
as well as the left-handed polarized electromagnetic waves with $\omega \sim \omega_{pe, loc}$ (where $\omega_{pe, loc}$ is the local plasma frequency).
Meanwhile there could be enhancement of the left-handed polarized quasiparallel electromagnetic
waves with $\omega \sim 2 \omega_{pe, loc}$ in the presence of both the forward ($L$) and backward ($L^{'}$) quasiparallel propagating
Langmuir waves, i.e., $L+L^{'} \rightarrow 2H_{p}$.
There are, indeed, energy enhancement of electromagnetic waves along the quasiparallel
directions with $\omega \sim \omega_{norm}$ as well as $\omega \sim 2 \omega_{norm}$ in both plasmas
(see column a1 in Figs.~\ref{Et_polarization_case_2} and \ref{Et_polarization_case_4}), whether the process of $L+L^{'} \rightarrow 2H_{p}$ contributes to
this energy enhancement, however, still needs further investigations in our future studies.
But it is certain that electromagnetic waves with $\omega \sim \omega_{norm}$ and $\omega \sim 2 \omega_{norm}$ along the quasiparallel directions do
not contribute to the energy peak of the left-handed polarized electromagnetic waves, which are located at $\omega \sim
\omega_{ce,mean}$, and propagating along the quasiperpendicular directions.

\section{Conclusions and Discussion}
\label{Conclusions}

In this paper we focus on studying the coherent emission mechanism behind the solar radio bursts.
Coherent solar radio emissions assuming homogeneous plasmas conditions at their source region have been widely studied via theoretical analysis or/and simulations.
There is, however, seldom plasma immersed in a uniform magnetic field in realistic solar coronal environment.
Inhomogeneities exist over almost all spatial scales of the solar corona.
As it is known, diverse macroscopic phenomena (e.g., convection, thermal conduction) can generate
inhomogeneity at large scales and many of them could cascade down to
kinetic scales, in particular via turbulence, which is ubiquitous in the solar corona.
Turbulence can also, in particular, be generated by processes
associated to solar flares and radio emission, like magnetic reconnection and shocks.
As mentioned by~\citealp{Melrose1975SoPh...43...79M} and~\citealp{Winglee&Dulk_1986ApJ...307..808W}, the inhomogeneous nature of
plasmas is quite important as it influences properties of the emission spectrum of radio bursts.
Contribution of the inhomogeneity of background magnetic field as well as density and/or temperature in plasmas
on the properties of the coherent radio emission for solar radio bursts, hence, needs to be further verified and expanded.
In this paper, we report the excitation properties of both the electromagnetic and the electrostatic waves produced by ring-beam distributed energetic electrons in
inhomogeneous magnetized equilibrium plasmas of solar corona via selfconsistent 2.5-dimensional particle-in-cell (PIC) code numerical simulations.
The disequilibrium introduced by the inhomogeneous background magnetic field is, we consider, balanced by either inhomogeneous density or inhomogeneous
temperature of the background plasma, corresponding to inhomogeneous-density plasma and inhomogeneous-temperature plasma, respectively.
And inhomogeneity of these plasmas exist only along the dimension perpendicular to the background magnetic field in this
study.

Same with homogeneous plasmas, the ring-beam distributed energetic electrons can
excite waves associated to the two most promising mechanisms for the coherent solar radio emission: the beam and ECM
instabilities. Those instabilities can excite electromagnetic waves for radio emissions.
These excitations occur due to the free energy (i.e., the positive gradients) along the parallel and perpendicular directions
to the ambient magnetic field in the energetic ring-beam electron velocity distribution, respectively.
However properties of these excited waves are not always the same between the inhomogeneous-density and inhomogeneous-temperature
plasmas. For instance,
\begin{description}
  \item[Similarities] The onset of the ECM instability is later than the beam instability.
                    The beam instability mainly excite electrostatic Langmuir waves propagating quasiparallel to the background magnetic field
                    as well as the electromagnetic whistler waves. The ECM instability, however, fully contributes to the excitation of electromagnetic
                    Z, O, and X-mode waves. Saturation energy of these excited electrostatic and electromagnetic waves have the same order of magnitude in these two
                    inhomogeneous plasmas.
                    For the electromagnetic waves, these stronger excited ones mainly propagate quasiperpendicular to the background magnetic
                    field, and most of their energies are located around $h \omega_{ce,mean}$, where $\omega_{ce,mean}$ is the mean value of the electron gyrofrequency
                    over the whole simulation domain and $h$ is the harmonic index.
                    The fundamental branch presents a left-handed polarization due to the energy dominance of the O-mode waves, while the X-mode waves determine the properties
                    of the second and third harmonic branches.
                    The low frequency whistler waves are more likely
                    found in region with a low plasma density and low temperature, which could reduce wave damping.
                    The backward quasiparallel Langmuir waves can be detected
                    after the excitation of the forward Langmuir waves.
  \item[Differences] \textbf{In the inhomogeneous-density plasma}, there is an excitation of the right-handed polarized
                    X1-mode waves contributed by the ECM instability. During the excitation period, most energy of the
                    X1-mode (whistler, Z, harmonics of the O, higher harmonics of the X-mode as well as the backward Langmuir) waves
                    are located around the region with a tenuous (dense) background plasma.
                    During the nonlinear period, most of these excited electromagnetic waves emplace their energy
                    more in the tenuous-plasma region, where the low plasma density could reduce the damping of these excited electromagnetic waves.
                    \textbf{In the inhomogeneous-temperature plasma}, excitation of the X1-mode wave is absent. Other excited electromagnetic waves
                    are more homogeneously distributed along the inhomogeneity gradient over the whole wave
                    evolution. There are also more energized backward Langmuir waves than those in the inhomogeneous-density plasma. These backward Langmuir waves could be
                    detected only around the region with a low temperature and a high density as the dense region of the inhomogeneous-density plasma.
                    Additionally, frequency bandwidth of different harmonic branches are thinner than those in the inhomogeneous-density plasma.
\end{description}

As prediction by those analytical investigations for the growth rate of different wave modes excited by the ECM
instability (e.g.,~\citealp{Tong_etal_2017PhPl...24e2902T, Chen_etal_2017JGRA..122...35C, Zhao_etal_2016ApJ...822...58Z}),
different source regions between the X1-mode and other electromagnetic modes in the inhomogeneous-density plasma are due to their excitations strongly
depending on the frequency ratio $\omega_{pe}/\omega_{ce}$. 
Temperature of the background plasma does, however, not influence much on the excitation of the
electromagnetic waves in non-relativistic space plasmas. Wave excitation properties in the inhomogeneous-temperature plasma are indeed similar to those in homogeneous
plasmas known from previous related investigations for homogeneous plasmas.

Based on the properties of the excited electromagnetic waves in this paper, our results
can be applied to explain some features of the solar radio bursts with zebra-stripe pattern,
e.g., lace burst~\citep{Karlicky_etal_2001A&A...375..638K},
fiber burst~\citep{Chernov_etal_2014Ge&Ae..54..406C},
zebra-pattern burst~\citep{Huang&Tan2012ApJ...745..186H, Chernov2015arXiv151206311C}
as well as narrowband spikes~\citep{Krucker_etal_1994A&A...285.1038K, Karlicky_etal_2021ApJ...910..108K, Karlicky_etal_2022SoPh..297...54K}:
\begin{description}
  \item[Slow frequency drifting rate] Most of the solar radio
                              bursts with zebra-stripe pattern often exhibit slow frequency drifting rates compared to those of
                              the solar type \uppercase\expandafter{\romannumeral3} radio
                              burst\citep{Aschwanden2005psci.book.....A, Tan_etal_2014ApJ...780..129T, Chernov2015arXiv151206311C}.
                              That indicates that source of these radio bursts might be slow electron beams.
                              Based on our results, the earlier state of the related slow electron beams could drift quite fast while they slow down
                              due to the onset of the beam instability. The beam instability could not always lead to efficient
                              wave excitation for radio emissions~\citep{Thurgood_2015A&A...584A..83T}.
                              Slow-down electron beams with population inversion along the direction perpendicular to the background magnetic filed
                              (e.g., for the ECM instability) are more likely to play an important role in the generation of those solar radio bursts
                              with slow frequency driftings.
  \item[Zebra-structure stripes] Our results show that, in dependence on the wave propagating angles $\theta$, one or more harmonic bands can escape from their source region,
                              which could potentially explain the detection of various number of harmonic branches in radio bursts with zebra-stripe
                              pattern.
                              Also, variation of frequency bandwidth in single stripe of radio bursts with zebra-stripe pattern
                              might be related to variation of the perpendicular plasma density gradient in the source region of
                              the stripes. Based on our results, a larger perpendicular density gradient in plasmas could lead to a wider frequency bandwidth
                              in each harmonic branch. Note that, according to the study of~\citealp{Yao_etal_2021JPlPh..87b9003Y},
                              parallel density gradient has little effects on the ECM processes.
                              Additionally, observed frequency separation between adjacent zebra-structure stripes are not always equidistant as well as
                              observed zebra-structure stripes
                              can be present at noninteger harmonics, which cannot be explained by a simple cyclotron harmonic emission.
                              Nonetheless, our results indicate that the Doppler effect due to the drifting motion of electron beams can influence
                              the frequency separation between adjacent stripes, and lead to noninteger harmonic emissions in particular when
                              their propagating angles $\theta$ are small.
\end{description}

Note that more widespread and standard mechanism for the solar radio bursts with zebra-stripe pattern are based on the double plasma resonance
(DPR) or Bernstein instabilities~\citep{Winglee&Dulk_1986ApJ...307..808W, Benacek&Karlicky_2018A&A...611A..60B, Karlicky_etal_2022SoPh..297...54K}.
Actually, the DPR, Bernstein, and ECM instabilities have the same free energy, i.e., a positive velocity gradient in the electron
distribution function perpendicular to the ambient magnetic field $\partial f / \partial v_{\bot} > 0$.
The main difference among these three instabilities is that the DPR and Bernstein (ECM) instability is applied to weakly (strongly) magnetized plasmas
with $\omega_{ce} < \omega_{pe}$ ($\omega_{ce} > \omega_{pe}$) and trigger excitation of the electrostatic
upper-hybrid and Bernstein waves (electromagnetic X- and O-mode waves), respectively.
How the electrostatic upper-hybrid and Bernstein waves transform into electromagnetic waves and are observed as radio
zebra-structure emissions should be considered for the DPR and Bernstein mechanism to explain the formation of the solar radio bursts
with zebra-stripe pattern~\citep{Li_etal_2021ApJ...909L...5L}.
The ECM instability can, however, excite electromagnetic waves directly.
The DPR and Bernstein model are more accredited for the emission mechanism of the solar radio bursts with zebra-stripe pattern since these models fulfill well
the standard plasma model of the solar corona with $\omega_{ce} < \omega_{pe}$~\citep{Wild_1985srph.book....3W}.
However it is still possible that $\omega_{ce} > \omega_{pe}$ exists in density
cavities of the solar corona due to ubiquitous Alfv\'{e}nic turbulence~\citep{Wu_etal_2014A&A...566A.138W, Chen_etal_2017JGRA..122...35C}
as well as in the low-density separatrices of magnetic reconnection~\citep{Drake_etal_2003Sci...299..873D, Pritchett&Coroniti_2004JGRA..109.1220P, Munoz2018PhRvE}.
Observations by~\citealp{Regnier_2015A&A...581A...9R, Morosan_etal_2016A&A...589L...8M}
have provided the existence of $\omega_{ce} > \omega_{pe}$ in the
solar corona above the core of active regions.
All above posts the ECM mechanism (in addition to the DPR and Bernstein models) as one of the valid candidates for the interpretation
of the solar radio bursts with zebra-stripe pattern.

Additionally, note that there still needs further investigations to discover whether the nonlinear wave-wave interactions
occur under the presence of the forward and backward Langmuir waves in these inhomogeneous plasmas.
But what we ascertain is that the nonlinear wave-wave interactions of the forward and backward Langmuir waves
do not contribute much to these most strongly excited electromagnetic waves, which are produced, for sure, by the ECM emission processes in this study.
In spite of the wave conversion efficiency of the plasma emission process, scattering of forward Langmuir waves to the backward Langmuir waves more likely
occurs in a cold and dense plasma. Most space plasmas, however, stay hot and tenuous in the solar corona.
The ECM emission processes are, hence, expected more for the generation of radio
bursts from the solar corona, where frequency requirement for a strong ECM
emission, i.e., $\omega_{ce}>\omega_{pe}$, could be easily satisfied in the
presence of Alfv\'{e}n waves~\citep{Wu_etal_2014A&A...566A.138W}.\\\\

The present research at PMO was supported by the Strategic Priority Research Program of the
Chinese Academy of Sciences under grant No.XDB0560000, the project of National Natural Science
Foundation of China No.12003073, 42174195, 11873018, and 11790302. And Jan Ben\'{a}\v{c}ek
acknowledges the support by the German Science Foundation (DFG) project BE 7886/2-1.
We gratefully acknowledge the developers of the ACRONYM code, the Verein zur F\"{o}rderung kinetischer Plasmasimulationen e.V.,
as we as the computing resources in the Max Planck Computing and Data Facility (MPCDF) at Garching, Germany, and the Max Planck Institute for Solar System
Research, Germany.





\end{document}